\documentclass[aps,prb,showpacs,preprint,
superscriptaddress,
]{revtex4-1}
\usepackage[graphicx]{realboxes}
\usepackage{dcolumn}
\usepackage{bm}
\usepackage{amssymb}
\usepackage{amsmath}
\usepackage{subfigure}
\usepackage{supertabular}
\usepackage{float}
\usepackage{graphicx}
\usepackage{dcolumn}
\usepackage{bm}
\usepackage{amsmath}
\usepackage{amsthm}
\usepackage{booktabs}
\usepackage{float}
\usepackage{braket}
\usepackage{color}
\usepackage{multirow}
\usepackage{siunitx}
\begin{document}

\preprint{}

\title{Decoupling Coherent and Particle-like Phonon Transport through Bonding Hierarchy in Soft Superionic Crystals}



\author{Wenjie Xiong}
\affiliation{College of Physics Science and Technology, Yangzhou University, Jiangsu 225009, China}

\author{Hao Huang}
\affiliation{Advanced Copper Industry College, Jiangxi University of Science and Technology, Yingtan 335000,China}

\author{Yu Wu}
\email{wuyu9573@qq.com}
\affiliation{Micro- and Nano-scale Thermal Measurement and Thermal Management Laboratory, School of Energy and Mechanical Engineering, Nanjing Normal University,Jiangsu,
Nanjing 210023, China}

\author{Xinji Xu}
\affiliation{College of Physics Science and Technology, Yangzhou University, Jiangsu 225009, China}


\author{Geng Li}
\affiliation{China Rare Earth Group Research Institute, Ganzhou, 341000, China}
\affiliation{Key Laboratory of Rare Earths, Ganjiang Innovation Academy, Chinese Academy of Sciences, Ganzhou, 341000, China}

\author{Zonglin Gu}
\email{guzonglin@yzu.edu.cn}
\affiliation{College of Physics Science and Technology, Yangzhou University, Jiangsu 225009, China}


\author{Shuming Zeng}
\email{zengsm@yzu.edu.cn}
\affiliation{College of Physics Science and Technology, Yangzhou University, Jiangsu 225009, China}

\date{\today}

\begin{abstract}
Within the framework of the unified theory thermal transport model, the competing contributions of coherent and incoherent terms create a trade-off relationship, posing substantial challenges to achieving a reduction in overall $\rm \kappa_L$. In this work, we theoretically demonstrate that the superionic crystals X$_6$Re$_6$S$_8$I$_8$ (X = Rb, Cs) exhibit ultralow glass-like and particle-like thermal conductivities. The weak interactions between free alkali metal ions X$^+$ (X = Rb, Cs) and I$^-$ anions induce pronounced lattice anharmonicity, which enhances phonon scattering and suppresses group velocities, thereby reducing the particle-like thermal conductivity ($\rm \kappa_p$). Concurrently, the significant bonding heterogeneity within the [Re$_6$S$_8$I$_6$]$^{4-}$ clusters promotes phonon dispersion flattening and low-frequency phonon localization. The resulting discretized phonon flat bands substantially diminish the glass-like thermal conductivity ($\rm \kappa_c$). At room temperature, the total $\rm \kappa_L$ of X$_6$Re$_6$S$_8$I$_8$ (X = Rb, Cs) falls below 0.2 Wm$^{-1}$K$^{-1}$. Furthermore, the bonding characteristics between X$^+$ and I$^{-1}$ anions induce an anomalous cation mass-independent stiffening of low-frequency phonon branches in this system, resulting in counterintuitive thermal transport behavior. This work elucidates fundamental mechanisms governing heat transfer in
ultralow $\rm \kappa_L$ materials and establishes novel pathways for transcending conventional thermal conductivity limitations.

\end{abstract}

\maketitle


\section{\label{intro}Introduction}
The exploration of materials with ultralow lattice thermal conductivity $\rm \kappa_L$ holds significant importance in both fundamental research and technological applications, such as thermal barrier coatings\cite{padture2002thermal} and thermoelectrics\cite{he2017advances}. Materials with intrinsically low $\rm \kappa_L$, typically governed by phonon-phonon scattering mechanisms, exhibit several distinctive features such as intrinsic rattling\cite{christensen2008avoided}, lattice anharmonicity\cite{qin2021power}, ferroelectric instability\cite{sarkar2020ferroelectric}, and structural complexity\cite{PhysRevLett.125.085901}. Based on these principles, a series of materials with low $\rm \kappa_L$ at room temperature such as Tl$_3$VSe$_4$ (0.30 Wm$^{-1}$K$^{-1}$)\cite{mukhopadhyay2018two}, TlSe (0.62 Wm$^{-1}$K$^{-1}$)\cite{dutta2019ultralow}, and Cu$_{12}$Sb$_4$S$_{13}$ (0.67 Wm$^{-1}$K$^{-1}$)\cite{PhysRevLett.125.085901} have been discovered. In addition, strategies like nanostructuring and entropy-driven point defect engineering can further suppress $\rm\kappa_L$\cite{jiang2021high}. The advent of machine learning methods has enabled efficient exploration of low  $\rm \kappa_L$ materials across vast chemical and structural spaces\cite{zeng2023ultralow}. However, the $\rm \kappa_L$ of crystalline materials cannot be arbitrarily low-it is fundamentally bounded\cite{zeng2024pushing}. To date, the experimentally measured $\rm \kappa_L$ of crystals at room temperature has not fallen below 0.2 Wm$^{-1}$K$^{-1}$, the search for materials with ultralow $\rm \kappa_L$ remains an active research frontier.

In the unified theory of lattice thermal conductivity proposed by Simoncelli et al.\cite{simoncelli2019unified,PhysRevX.12.041011}, heat transport is governed by dual contributions from localized diffusons (glass-like thermal conductiviy $\rm \kappa_c$) and propagating phonons (particle-like thermal conductivity $\rm \kappa_p$). Generally, in simple lattices with few atoms per primitive cell (PC), $\rm \kappa_L$ is dominated by propagating phonons. However, in complex crystals with many atoms per PC, the contribution of localized diffusons becomes significant and cannot be neglected. For example, at 300 K, PbTe (2 atoms per PC) exhibits a total $\rm \kappa_L$ of ~2 Wm$^{-1}$K$^{-1}$, which is almost entirely contributed by $\rm \kappa_p$\cite{PhysRevB.85.184303}. In contrast, the complex crystal Ag$_8$GeS$_6$ (60 atoms per PC)
demonstrates an ultralow $\rm \kappa_p$ of merely 0.04 Wm$^{-1}$K$^{-1}$, which comparable to that of air, while its $\rm \kappa_c$ reaches 0.43 Wm$^{-1}$K$^{-1}$, significantly exceeding $\rm \kappa_p$\cite{zeng2024pushing}. A viable strategy for discovering materials with ultralow $\rm \kappa_L$ involves either minimizing $\rm \kappa_p$ in simple lattices or suppressing $\rm \kappa_c$ in complex crystal structures. For the former case, Zeng et al. identified AgTlI$_2$ (I4/mcm) as a rare simple crystalline system simultaneously exhibiting low $\rm \kappa_p$ and suppressed $\rm \kappa_c$\cite{zeng2024pushing}, achieving an unprecedented room-temperature $\rm \kappa_L$ of 0.25 Wm$^{-1}$K$^{-1}$.
For the latter case (complex crystals), whether materials with intrinsically low $\rm \kappa_c$ exist remains an open fundamental question in thermal transport physics.

In this work, we systematically investigate the lattice thermal conductivity of the recently discovered superionic compounds\cite{kamaya2011lithium, wang2015design,liu2012copper} X$_6$Re$_6$S$_8$I$_8$ (X = Rb, Cs)\cite{sun20210d} using first-principles calculations combined with the temperature-dependent effective potential (TDEP) method. By rigorously treating three- and four-phonon scattering processes within the unified theory of lattice thermal transport, we reveal exceptionally suppressed contributions from both $\rm \kappa_p$ and $\rm \kappa_c$,
leading to an ultralow $\rm \kappa_L$ below 0.2 Wm$^{-1}$K$^{-1}$. The quasi zero-dimensional (0D) structure of X$_6$Re$_6$S$_8$I$_8$ consists of isolated [Re$_6$S$_8$I$_6$]$^{4-}$ clusters embedded in a three-dimensional ionic framework formed by Rb$^+$/Cs$^+$ cations and bridging I$^-$
anions.  Strong phonon anharmonicity and localization in this architecture generate extensive optical flat bands that suppress both particle-like and wave-like thermal transport. Counterintuitively, despite the heavier atomic mass of Cs-which typically reduces phonon group velocities ($v_{\mathrm{g}}$) -our calculations reveal higher thermal conductivity in Cs-based compounds than in their Rb analogs. This anomaly arises from distinct X-I interactions that reshape local lattice dynamics and bonding, unexpectedly enhancing phonon lifetimes and propagation in Cs variants. These results defy the conventional``liquid-like thermal conduction" model and establish new principles for engineering ultralow $\rm \kappa_L$ materials.

\section{\label{intro}COMPUTATIONAL METHODS}
All first-principles calculations were performed using the Vienna \textit{Ab Initio} Simulation Package (VASP) \cite{kresse1996efficient}, based on density functional theory (DFT)\cite{hohenberg1964density,kohn1965self}. The projector augmented wave (PAW) method \cite{kresse1999ultrasoft} with a plane-wave basis set was employed to describe ion-electron interactions, while exchange-correlation effects were treated within the revised Perdew-Burke-Ernzerhof functional optimized for solids (PBEsol) \cite{perdew1996generalized}.
A plane-wave energy cutoff of 520 eV was imposed throughout the calculations. Structural relaxations utilized a 6 $\times$ 6 $\times$ 6 Monkhorst-Pack \textbf{k}-mesh for Brillouin zone sampling, with
convergence thresholds of 10$^{-8}$ eV for energy and 10$^{-4}$ eV/\AA\ for atomic forces. Born effective charge tensors and dielectric constants were computed via density functional perturbation theory (DFPT)\cite{baroni2001phonons}, incorporating nonanalytic corrections to the dynamical matrix. To determine interatomic force constants (IFCs), we conducted \textit{ab initio} molecular dynamics (AIMD) simulations in the NVT ensemble using a 2$\times$2$\times$2 supercell containing 224 atoms. The
system was equilibrated for 20 ps with a timestep of 1 fs. Temperature-dependent effective potential (TDEP) methodology \cite{hellman2013temperature} was subsequently applied to extract second-, third-,
and fourth-order IFCs at varying temperatures, employing cutoff radii of 8 \AA, 6 \AA, and 4 \AA\ respectively. These cutoff values ensured comprehensive inclusion of atomic interactions governing the material's
dynamical response.

Within the Wigner formalism, the lattice thermal conductivity tensor $\rm \kappa^{\alpha\beta}_L$ decomposes into two contributions:
\begin{equation}
\rm{\kappa^{\alpha\beta}_L} = \rm{\kappa^{\alpha\beta}_p} + \rm{\kappa^{\alpha\beta}_c},
\end{equation}
where the particle-like term $\rm \kappa^{\alpha\beta}_p$ describes semiclassical phonon transport:
\begin{equation}
	\rm{\kappa^{\alpha\beta}_p} = \frac{1}{N_q V} \sum_{qs} C^s_q \nu^s_{q,\alpha} \nu^s_{q,\beta} \tau^s_q.
\end{equation}
Here, $C^s_q = \hbar\omega^s_q \partial n^s_q/\partial T$ denotes the mode-resolved heat capacity, with $n^s_q$ being the Bose--Einstein distribution. The coherence contribution $\rm \kappa^{\alpha\beta}_c$ captures wave-like tunneling effects between distinct phonon branches $s$ and $s'$:
\begin{equation}
	\kappa^{\alpha\beta}_c = \frac{\hbar^2}{k_B T^2 V N_q} \sum_q \sum_{s \neq s'} \frac{\omega^s_q + \omega^{s'}_q}{2} \nu^{s,s'}_{q,\alpha} \nu^{s,s'}_{q,\beta} \times \frac{\omega^s_q n^s_q(n^s_q + 1) + \omega^{s'}_q n^{s'}_q(n^{s'}_q + 1)}{4(\omega^s_q - \omega^{s'}_q)^2 + (\Gamma^s_q + \Gamma^{s'}_q)^2} (\Gamma^s_q + \Gamma^{s'}_q)
  \end{equation}
where $\nu^{s,s'}_{q,\alpha}$ represents the interband velocity matrix element. The Cartesian indices $\alpha,\beta \in \{x,y,z\}$ span spatial dimensions, $V$ is the unit cell volume, and $N_q$ enumerates sampled phonon wavevectors in the Brillouin zone. Scattering rates $\Gamma^s_q$ quantify phonon lifetime broadening. This dual formulation reconciles semiclassical transport ($\rm \kappa_p$) with quantum coherence effects ($\rm \kappa_c$), particularly
significant in materials exhibiting phonon branch degeneracies or strong anharmonicity. The $\rm \kappa^{\alpha\beta}_L$ was computed using an in-house modified version of the FourPhonon code\cite{li2012thermal,li2014shengbte,han2022fourphonon, PhysRevB.109.214307}, incorporating three- and four-phonon scattering processes, and performed in a $q$-mesh of $10 \times 10 \times 10$.

\section{\label{result}Results and discussions}

\begin{figure}[ht]
	\centering
	\includegraphics[width=\textwidth,height=\textheight,keepaspectratio]{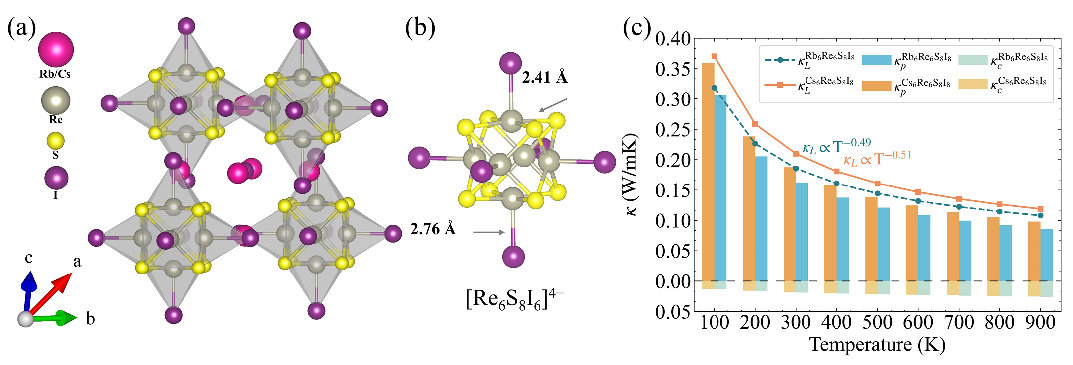}
	\caption{(a)The crystal structure of X$_6$Re$_6$S$_8$I$_8$ (X = Rb, Cs); (b) The Isolated [Re$_6$S$_8$I$_6$]$^{4-}$ cluster; (c) The total lattice thermal conductivities ($\rm \kappa_{L}$), the particle-like lattice thermal conductivities ($\rm \kappa_p$) and the glass-like lattice thermal conductivities ($\rm \kappa_c$) of X$_6$Re$_6$S$_8$I$_8$ (X = Rb, Cs) from 100K to 900K.}
	\label{fig:figure1}
\end{figure}

The superionic crystals X$_6$Re$_6$S$_8$I$_8$ (X = Rb, Cs) adopt a cubic 0D perovskite structure \cite{sun20210d}, with Rb$_6$Re$_6$S$_8$I$_8$ having been experimentally synthesized \cite{laing2024solution}. As shown in Fig.~\ref{fig:figure1}(a), X$_6$Re$_6$S$_8$I$_8$ crystallizes in the space group \textit{Fm$\overline{3}$m}, comprising alkali metal ions (X$^+$), iodide ions (I$^-$), and discrete [Re$_6$S$_8$I$_6$]$^{4-}$ clusters. The [Re$_6$S$_8$I$_6$]$^{4-}$ cluster (Fig.~\ref{fig:figure1}(b)) serves as the fundamental building unit, featuring an octahedral
Re$_6$ core stabilized by direct metal-metal bonds (Re-Re: 2.60~\SI{}{\angstrom}). Each face of the Re$_6$ octahedron is capped by a sulfur atom in $\mu_3$ coordination (Re-S: 2.41~\SI{}{\angstrom}), forming a [Re$_6$S$_8$]$^{2+}$ core. Peripheral iodides coordinate to the octahedron through Re-I bonds (2.76~\SI{}{\angstrom}), completing the [Re$_6$S$_8$I$_6$]$^{4-}$ cluster. Alkali metal atoms (X) occupy interstitial sites between clusters, exhibiting weak X-I interactions (Rb-I: 3.94~\SI{}{\angstrom}) that stabilize the crystal framework. Additional X-I-X bridging
motifs (Rb-I: 3.88~\SI{}{\angstrom}) fill structural voids, forming ionic coordination networks. Optimized structural parameters for Rb$_6$Re$_6$S$_8$I$_8$ (Re-Re: 2.60~\SI{}{\angstrom}, Re-S: 2.40~\SI{}{\angstrom}, Re-I: 2.77~\SI{}{\angstrom}) show agreement with experimental data \cite{laing2024solution}. Detailed structural parameters for both Rb and Cs analogs are provided in Table S1 (see supplementary information). The system exemplifies a rattling model, where X atoms (Rb/Cs) behave as loosely bound "rattler" guests within the rigid [Re$_6$S$_8$I$_6$]$^{4-}$ host framework.

The electronic structures of X$_6$Re$_6$S$_8$I$_8$ are presented in Fig.~S1. Both Rb$_6$Re$_6$S$_8$I$_8$ and Cs$_6$Re$_6$S$_8$I$_8$ exhibit indirect bandgap semiconductor behavior, with valence band maxima (VBM) at the $\Gamma$-point and conduction band minima (CBM) at the $L$-point, yielding bandgaps ($E_g$) of 2.42~eV and 2.45~eV, respectively. Atomic projected electronic density of states (EDOS) analysis reveals dominant contributions from S and I atomic orbitals near the VBM, while Re atomic orbitals primarily constitute the CBM.
To assess thermal stability, we performed 20~ps NVT-ensemble AIMD simulations at varying temperatures (see Fig.~S2). Remarkably, both compounds maintain structural integrity up to 800~K, evidenced by minimal fluctuations in total energy. Phonon dispersion calculations within the harmonic approximation confirm dynamical stability, showing no imaginary frequencies across the Brillouin zone (Fig.~S3). This stability persists under finite-temperature conditions as validated by TDEP calculations up to 800~K. Notably, temperature evolution induces distinct phonon renormalization: mid-to-high-frequency optical branches soften with increasing temperature, whereas low-frequency optical modes exhibit hardening behavior. This anomalous temperature dependence correlates with atomic vibration patterns -- mid/high-frequency modes predominantly involve Re-S framework vibrations, while low-frequency modes arise from I-Rb/I-Cs interactions. The hardening of low-frequency optical branches suggests enhanced anharmonic coupling between Rb$^+$ and I$^-$ ionic bonds at elevated temperatures. 

The cage-like crystal structure of X$_6$Re$_6$S$_8$I$_8$ (X = Rb, Cs) engenders distinctive thermal transport characteristics\cite{chang2018anharmoncity}. Figure~\ref{fig:figure1}(c) presents the temperature-dependent $\rm \kappa_L$ ($\rm \kappa_L$ = $\rm \kappa_p$ + $\rm \kappa_c$) decomposition, revealing particle-like $\rm \kappa_p$ and coherent $\rm \kappa_c$ contributions. At 300~K, Cs$_6$Re$_6$S$_8$I$_8$ and Rb$_6$Re$_6$S$_8$I$_8$ exhibit ultralow $\kappa_{\mathrm{L}}$ values of 0.19 and 0.17~W\,m$^{-1}$\,K$^{-1}$, respectively, surpassing recently reported ultralow
$\kappa_{\mathrm{L}}$ materials including Ag$_8$GeTe$_6$ and AgTlI$_2$, as well as conventional thermoelectric materials (see Table~\ref{tbl:table1}). Notably, both compounds demonstrate weak temperature dependence with $\rm {\kappa_L}$ $\propto T^{-0.51}$ (Cs) and $T^{-0.49}$ (Rb),
signaling strong lattice anharmonicity. The particle contribution dominates the thermal transport, accounting for 90.2\% (Cs) and 87.9\% (Rb) of $\rm \kappa_L$ at 300~K. This hierarchy suggests that
diffusive phonon transport governs heat conduction in these superionic crystals. Intriguingly, the heavier Cs analog exhibits higher $\rm \kappa_L$ than its Rb counterpart -- a mass dependence reversal contradicting conventional Slack formalism predictions \cite{slack1973nonmetallic,slack1979thermal}. Structural analysis (SI Table~1) reveals significant lattice parameter and bond length variations between the analogs. These structural distinctions likely modify interatomic force constants, thereby altering phonon scattering phase space.

\begin{table*}[ht]
	\centering
	\caption{The elastic constants (C$_{ij}$), the bulk modulus (B), shear modulus (G), Young's modulus (Y), and average wave velocity ($\upsilon$) of X$_6$Re$_6$S$_8$I$_8$, unit in GPa.}
	\label{tbl:table1}
	\small
	\begin{tabular*}{\textwidth}{@{\extracolsep{\fill}}ccccccc@{}}
	\toprule
	\noalign{\smallskip}\hline
	Material & $C_{11}$ & $C_{12}$ & $C_{44}$ & $B$ & $G$ & $E$ \\
	\midrule
	\noalign{\smallskip}\hline
	Rb$_6$Re$_6$S$_8$I$_8$ & 14.54 & 7.75 & 5.36 & 10.01 & 4.58 & 11.91 \\
	Cs$_6$Re$_6$S$_8$I$_8$ & 16.69 & 5.55 & 6.11 & 9.27  & 5.90 & 14.59 \\
	\bottomrule
	\noalign{\smallskip}\hline
    \end{tabular*}
\end{table*}

For cubic symmetry systems, the independent elastic tensor components reduce to $C_{11}$, $C_{12}$, and $C_{44}$, as presented in Table~\ref{tbl:table1}. The elastic anisotropy factor ($A$), characterizing mechanical property variations along different crystallographic directions, is defined as: $A = \frac{2C_{44}}{C_{11} - C_{12}}$. Our calculations yield $A = 1.58$ for Rb$_6$Re$_6$S$_8$I$_8$ and $A = 1.10$ for Cs$_6$Re$_6$S$_8$I$_8$, both approaching unity, indicative of quasi-isotropic mechanical behavior. Notably, the larger anisotropy in Rb$_6$Re$_6$S$_8$I$_8$
suggests directional constraints in phonon propagation that may enhance scattering processes. Comparative analysis of elastic moduli reveals weaker interatomic bonding and a more loosely packed structure in Rb$_6$Re$_6$S$_8$I$_8$, as evidenced by its lower elastic modulus. This structural characteristic correlates with enhanced lattice anharmonicity, which restricts phonon transport velocities . Supporting this interpretation, Table~\ref{tbl:table2} demonstrates substantially reduced sound velocities in Rb$_6$Re$_6$S$_8$I$_8$ compared to its cesium counterpart. Significantly, the X$_6$Re$_6$S$_8$I$_8$ system exhibits a remarkably low $\rm \kappa_L$ comparable to  traditional thermoelectric materials, suggesting promising potential for energy conversion applications.

\begin{table*}[ht]
	\centering
	\caption{Comparative sound velocities ($\upsilon$) and lattice thermal conductivities at 300~K}
	\label{tbl:table2}
	\small
	\begin{tabular*}{\textwidth}{@{\extracolsep{\fill}}lcc@{}}
	\toprule
	\noalign{\smallskip}\hline
	Material & Sound velocity $\upsilon$ (m~s$^{-1}$) & $\kappa_L$ (W~m$^{-1}$K$^{-1}$) \\
	\midrule
	\noalign{\smallskip}\hline
	\textbf{Rb$_6$Re$_6$S$_8$I$_8$} (this work) & 1040.78 & 0.17 \\
	Ag$_8$GeTe$_6$ & 1099 & 0.25 \\
	\textbf{Cs$_6$Re$_6$S$_8$I$_8$} (this work) & 1163.06 & 0.19 \\
	Ag$_8$SnSe$_6$$^\dagger$ & 1494 & 0.36 \\
	Cu$_8$GeSe$_6$$^\dagger$ & 1793 & 0.3 \\
	PbSe$^\dagger$ & 1805 & 1.5 \\
	PbTe$^\dagger$ & 1814 & 1.9 \\
	\bottomrule
	\noalign{\smallskip}\hline
	\end{tabular*}
	\vspace{0.2cm}
	\raggedright\footnotesize $^\dagger$ Values from literature references\cite{li2016low,zhu2010improved}
\end{table*}

\begin{figure}[ht]
	\centering
	\includegraphics[width=\textwidth,height=\textheight,keepaspectratio]{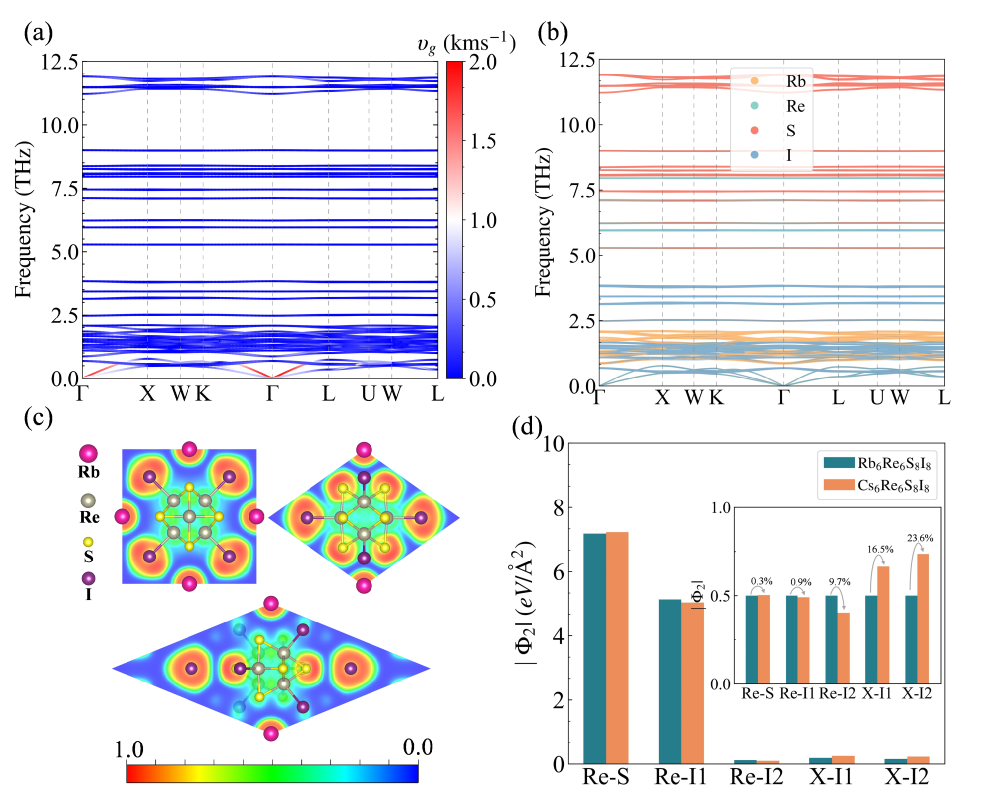}
	\caption{(a)The phonon dispersion of Rb$_6$Re$_6$S$_8$I$_8$ at 300 K with group velocities $v_{\mathrm{g}}$ projection; (b) The projected phonon dispersion with atomic projection of Rb$_6$Re$_6$S$_8$I$_8$ at 300 K; (c) The electronic localization function (ELF) corresponding to the (0,1,1), (1,2,1), and (-1,2,-1) planes of Rb$_6$Re$_6$S$_8$I$_8$; (d) The norm of the second-order force constants $\lvert \Phi_2 \rvert$ for the top three strongest bonds in X$_6$Re$_6$S$_8$I$_8$ (X = Rb, Cs). The center is the results of the relative value of the norm of the $\lvert \Phi_2 \rvert$ in X$_6$Re$_6$S$_8$I$_8$ (X = Rb, Cs).}
	\label{fig:figure2}
\end{figure}

To elucidate the origin of the low lattice thermal conductivity $\kappa_{\mathrm{L}}$ in X$_{6}$Re$_{6}$S$_{8}$I$_{8}$ compounds, we take Rb$_{6}$Re$_{6}$S$_{8}$I$_{8}$ as a prototype system. Figure~\ref{fig:figure2}(a) displays the projected phonon group velocities $v_{\mathrm{g}}$ along the phonon dispersion at 300~K. The phonon dispersion exhibits remarkable flatness across the entire frequency spectrum, particularly in optical branches, indicating spatially localized vibrational modes that hinder efficient energy propagation. All phonon modes demonstrate exceptionally low $v_{\mathrm{g}}$, with maximum values below 2~km\,s$^{-1}$ and majority below 0.5~km\,s$^{-1}$, signifying sluggish heat transfer kinetics. The projected phonon eigenvectors in Fig.~\ref{fig:figure2}(b) reveal distinct vibrational characteristics: acoustic branches primarily involve I atoms, while low-frequency optical branches (\textless 2.5~THz) arise from coupled vibrations of Rb and I atoms. Notably, these low-frequency modes (\textless2.5~THz) dominate thermal transport, suggesting significant contributions from Rb--I vibrations to $\kappa_{\mathrm{L}}$. Both phonon dispersion and group velocity characteristics correlate strongly with interatomic bonding environments. Figure~\ref{fig:figure2}(c) and Fig.S4 present the electron localization function (ELF) analysis, complemented by Bader charge calculations in Table~\ref{tbl:table3}. Clearly, two distinct iodine species emerge: (1) Covalent I1 within the [Re$_{6}$S$_{8}$I$_{6}$]$^{4-}$ octahedral clusters shows moderate electron localization (ELF $\approx$ 0.5) with incomplete charge transfer from Re, and (2) Ionic I exhibits negligible electron localization (ELF $\approx$ 0). The [Re$_{6}$S$_{8}$I$_{6}$]$^{4-}$ clusters maintain partial covalent bonding (Re--S/I ELF $\approx$ 0.5), while Rb atoms demonstrate complete electron transfer (ELF $\approx$ 0) characteristic of ionic bonding. This dual bonding architecture -- stable covalent
clusters interspersed with ionic Rb$^{+}$ and I$^{-}$ species -- creates weakly bonded structural frameworks. The resulting soft inter-cluster interactions promote extensive phonon band flattening, particularly in optical branches, ultimately suppressing $\kappa_{\mathrm{L}}$ through enhanced phonon scattering and reduced phonon group velocities.

\begin{table*}[ht]
    \centering
    \small
    \caption{Bader effective charges of constituent atoms in X$_6$Re$_6$S$_8$I$_8$ (X = Rb, Cs), unit in e$^-$.}
    \label{tbl:table3}
    \begin{tabular}{lccccc}
        \toprule
		\noalign{\smallskip}\hline
        Element & Formal charge & \multicolumn{2}{c}{Bader effective charge} \\
        \cmidrule(lr){3-4}
        & & X = Rb & X = Cs \\
        \midrule
		\noalign{\smallskip}\hline
        Rb/Cs  & +1 & 0.836 & 0.999 \\
        Re      & +3 & 0.784 & 0.603 \\
        S       & -2 & -0.668 & -0.648 \\
        I$_{1}$ & -1 & -0.480 & -0.456 \\
        I$_{2}$ & -1 & -0.751 & -0.850 \\
        \bottomrule
		\noalign{\smallskip}\hline
    \end{tabular}
\end{table*}

\begin{figure}[ht]
	\centering
	\includegraphics[width=\textwidth,height=\textheight,keepaspectratio]{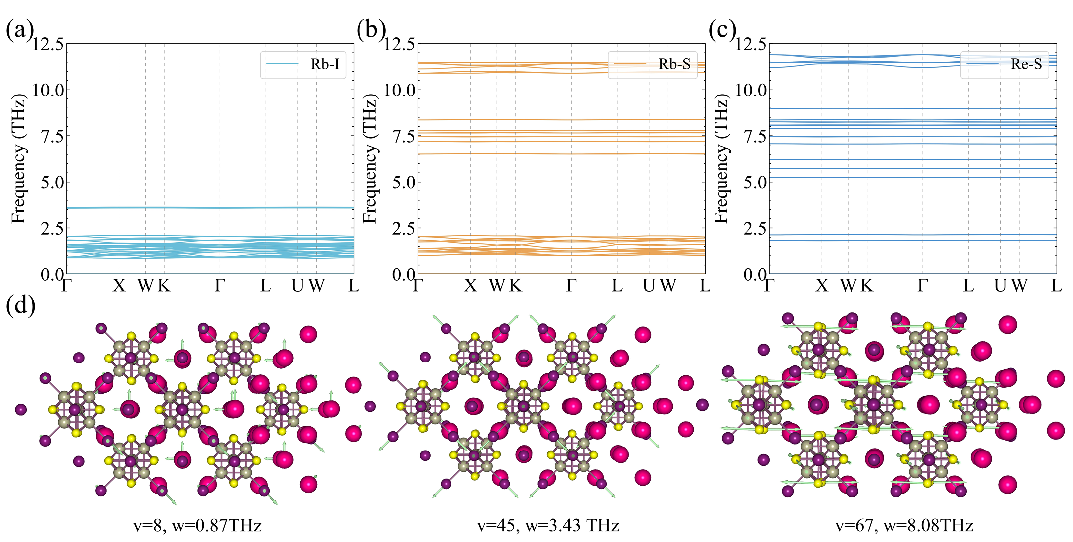}
	\caption{(a) The phonon dispersion between Rb and I, (b) between Rb and S, and (c)between Re and S for Rb$_6$Re$_6$S$_8$I$_8$ at 300 K. The visualization of phonon mode at point in (e) low-frequency {$\omega$=0.87 THz}, (f) low-medium-frequency {$\omega$=3.47 THz}, and (g) high-frequency {$\omega$=8.08 THz} optical branches.v is the v-th phonon branch.}
	\label{fig:figure3}
\end{figure}

The strength of interatomic interactions can be characterized through force constants. Figure~\ref{fig:figure2}(d) displays the second-order average interaction force constants ($\lvert \Phi_2 \rvert$) between key atomic pairs. Both Re--S and Re--I1 pairs exhibit large $\lvert \Phi_2 \rvert$, confirming the formation of stable [Re$_6$S$_8$I$_6$]$^{4-}$ octahedral clusters. In contrast, the weak interactions between alkali metal cations X$^{+}$ and ionic I$^{-}$ suggest that X$_6$Re$_6$S$_8$I$_8$ can be viewed as mobile X$^{+}$ cations and ionic I$_{2}^{-}$ species within the rigid [Re$_6$S$_8$I$_6$]$^{4-}$ framework. To explicitly demonstrate the origin of phonon band flattening, we reconstructed phonon spectra by selectively retaining specific interatomic force constants, as shown in Fig.~\ref{fig:figure3}(a-c). Notably, all retained interactions produce exceptionally flat phonon dispersions except for Re--S-derived optical branches above 10~THz, where stronger bonding induces greater dispersion. Figure~\ref{fig:figure3}(d) illustrates characteristic optical phonon modes across frequency domains at $\Gamma$ point. The low-frequency modes primarily correspond to the relative vibrations between Rb and I, the mid-frequency modes are attributed to the vibrations of I relative to the [Re$_6$S$_8$I$_6$]$^{4-}$ cluster, and the high-frequency modes arise from the internal atomic vibrations within the [Re$_6$S$_8$I$_6$]$^{4-}$ cluster. This mode localization hierarchy directly correlates with the hierarchy of bonding strengths revealed by $\lvert \Phi_2 \rvert$ analysis.

\begin{figure}[H]
	\centering
	\includegraphics[width=0.8\textwidth]{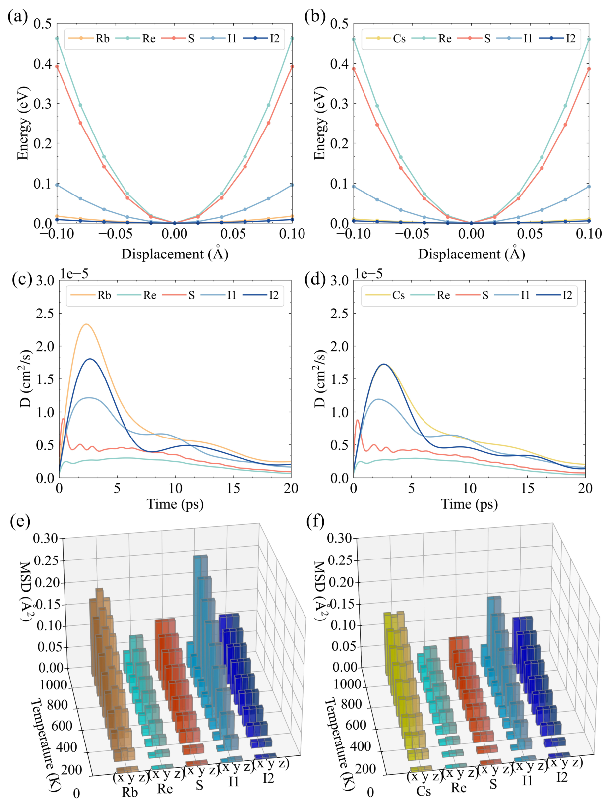}
	\caption{The potential energy as a function of the atomic displacement of (a) Rb$_6$Re$_6$S$_8$I$_8$, and (b) Cs$_6$Re$_6$S$_8$I$_8$ along the a-direction. The diffusion coefficient of (c) Rb$_6$Re$_6$S$_8$I$_8$, and (d) Cs$_6$Re$_6$S$_8$I$_8$ (d) at 300 K. The temperature-dependent atomic mean square displacements (MSD) of (e) Rb$_6$Re$_6$S$_8$I$_8$, and (f) Cs$_6$Re$_6$S$_8$I$_8$ (f).}
	\label{fig:figure4}
\end{figure}

Figures~\ref{fig:figure4}(a,b) presents the potential energy curves for atomic displacements along the $a$-axis in X$_6$Re$_6$S$_8$I$_8$(X = Rb, Cs). The steepest potential energy curves correspond to Re and S atoms, indicating strong resistance to displacement from their local environments. In contrast, remarkably flat potential energy curves are observed for X and I2 atoms, suggesting their high mobility within the lattice. The overall lattice anharmonicity primarily stems from the motion of X and I atoms. Notably, the potential energy landscapes surrounding each atom type in Rb$_6$Re$_6$S$_8$I$_8$ and Cs$_6$Re$_6$S$_8$I$_8$ exhibit nearly identical chemical environments. Figures~\ref{fig:figure4}(c,d) display the diffusion coefficients at 300 K, revealing that X and I atoms exhibit diffusion coefficients ($D$) more than twice as large as those of Re and S atoms. This further confirms the loosely bound nature of Rb and I atoms moving between [Re$_6$S$_8$I$_6$]$^{4-}$ clusters, while Re and S atoms remain constrained within the clusters. The temperature-dependent mean square displacements (MSDs) are shown in Figs.~\ref{fig:figure4}(e,f). All atoms demonstrate increased MSDs at higher temperatures, with X and I2 atoms showing particularly large displacements in $x$, $y$, and $z$ directions. Interestingly, I1 exhibits anisotropic behavior - small MSD along $x$-direction but large along $y$ and $z$-directions, reflecting distinct chemical environments. Compared with Cs$_6$Re$_6$S$_8$I$_8$, the Rb$_6$Re$_6$S$_8$I$_8$ compound demonstrates greater temperature-induced variations in MSDs, consistent with its softer lattice dynamics as evidenced by sound velocity and $\lvert \Phi_2 \rvert$ analysis.

\begin{figure}[ht]
      \centering
      \includegraphics[width=\textwidth,height=\textheight,keepaspectratio]{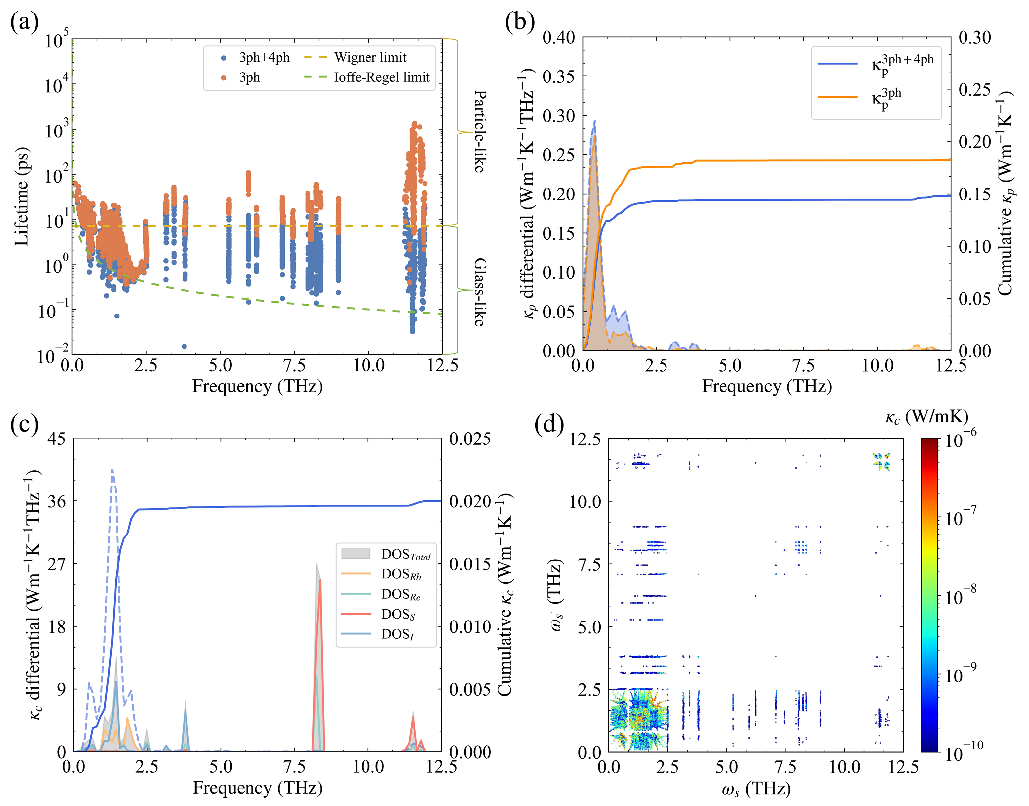}
      \caption{(a) The phonon lifetimes of the 3ph and 4ph as a function of phonon frequencies for Rb$_6$Re$_6$S$_8$I$_8$ at 300 K. Where the Ioffe-Regel ($\rm \tau_{Ioffe-Regel} = \frac{1}{\omega}$) and Wigner ($\rm \tau_{Wigner} = \frac{1}{\Delta \omega_{avg}}$) limits are represented by dotted yellow lines and green lines. (b) Calculated cumulative and differential $\rm \kappa_p$ as a function of phonon frequencies for Rb$_6$Re$_6$S$_8$I$_8$ at 300 K.(c) Calculated cumulative and differential $\rm \kappa_c$ as a function of phonon frequencies for X$_6$Re$_6$S$_8$I$_8$ at 300 K. (d) The resolved $\rm \kappa_c$ associated with various pairs of phonon frequencies ($\rm \omega_s$ and $\rm \omega_{s'}$).}
       \label{fig:figure5}
       \end{figure}

As shown in Fig.~\ref{fig:figure5}(b) and Fig.~S6(b), the calculated differential and cumulative $\rm \kappa_p$ of X$_6$Re$_6$S$_8$I$_8$ (X = Rb, Cs) at 300~K is derived using temperature-dependent force constants. The particle-like thermal conductivity component $\kappa_{\mathrm{p}}^{\mathrm{3ph}}$ for Rb$_6$Re$_6$S$_8$I$_8$ and Cs$_6$Re$_6$S$_8$I$_8$ under pure three-phonon (3ph) scattering are 0.17 and 0.18~W\,m$^{-1}$\,K$^{-1}$, respectively. Upon inclusion of four-phonon (4ph) scattering, the $\kappa_{\mathrm{p}}^{\mathrm{3ph+4ph}}$ values
decrease to 0.15 and 0.17~~W\,m$^{-1}$\,K$^{-1}$, with the Rb-analogue exhibiting a reduction exceeding 10\%, indicating the non-negligible contribution of four-phonon scattering. The $\kappa_{\mathrm{p}}^{\mathrm{3ph+4ph}}$ values of X$_6$Re$_6$S$_8$I$_8$ are significantly lower than those of other superionic crystals, e.g., the reported $\kappa_{\mathrm{p}}^{\mathrm{3ph+4ph}}$ is 0.32 ~W\,m$^{-1}$\,K$^{-1}$ for Re$_6$Se$_8$Cl$_2$\cite{zeng2025strain}. This reduction can be attributed to the structural complexity of the primitive cell. Specifically, Rb$_6$Re$_6$S$_8$I$_8$ exhibits a more complex primitive cell containing 28 atoms \cite{xia2023unified}, compared to only 16 atoms in Re$_6$Se$_8$Cl$_2$. The $\kappa_{\mathrm{p}}$ generally decreases with increasing number of atoms per
primitive cell, as exemplified by Ag$_9$GaS$_6$ containing 56 atoms per primitive cell, which exhibits an ultralow $\kappa_{\mathrm{p}} = 0.031$~W$\cdot$m$^{-1}$$\cdot$K$^{-1}$ \cite{PhysRevB.111.064307}. This negative correlation arises from enhanced phonon scattering channels in systems with larger primitive cells.

Figure~\ref{fig:figure5}(b) reveals two distinct growth regions of $\kappa_{\mathrm{p}}^{\mathrm{3ph+4ph}}$ in Rb$_6$Re$_6$S$_8$I$_8$: a low-frequency regime (0--0.8~THz) contributing 83\% of the total thermal conductivity, dominated by acoustic branches and the three lowest-frequency optical branches, followed by a low-frequency regime (0.8--1.7~THz) contributing 11\% from low-lying optical modes. A similar frequency dependence is observed in Cs$_6$Re$_6$S$_8$I$_8$, where phonons below 2.5~THz account
for 95\% of $\kappa_{\mathrm{p}}^{\mathrm{3ph+4ph}}$. The phonon lifetime ($\tau$) distributions at 300~K, presented in Fig.~\ref{fig:figure5}(a) and Supplementary Fig.~S6(a), demonstrate significant four-phonon scattering effects. Most phonons exhibit lifetimes below 10~ps, coupled with low group velocities, collectively suppressing $\kappa_{\mathrm{p}}$. In contrast, low-frequency phonons maintain extended lifetimes ($>$10~ps) and higher group velocities, leading to their dominant contributions to $\kappa_{\mathrm{p}}$.

Within the framework of the unified lattice thermal transport theory\cite{simoncelli2019unified,PhysRevX.12.041011}, the structural complexity of crystals inherently generates substantial off-diagonal thermal conductivity components $\kappa_{\mathrm{c}}$ arising from
wave-like phonon tunneling. As shown in Figs.~\ref{fig:figure5}(a) and S6(a), phonon modes with lifetimes exceeding the Wigner limit ($\tau > \Delta{\omega}_{avg}^{-1}$) predominantly contribute to the
particle-like thermal conductivity $\kappa_{\mathrm{p}}$, while those with lifetimes between the Wigner and Ioffe-Regel limits ($\omega^{-1} < \tau < \Delta{\omega}_{avg}^{-1}$) primarily
govern the wave-like tunneling component $\kappa_{\mathrm{c}}$. Here, $\rm \Delta \omega_{avg}$ is the average phonon band spacing, defined as $\rm \Delta \omega_{avg} = \frac{\omega_{max}}{3N}$, $\rm \omega_{max}$ is the highest phonon frequency,
and N is the number of atoms in the primitive cell. In the inclusion of 4ph scattering significantly reduces phonon lifetimes in Rb$_6$Re$_6$S$_8$I$_8$, driving numerous optical branches into the wave-like tunneling regime, which results in a pronounced reduction in
$\kappa_{\mathrm{p}}^{\mathrm{3ph+4ph}}$. Despite the abundance of phonon modes in the wave-like tunneling regime, their contributions to $\kappa_{\mathrm{c}}$ remain negligible due to the predominance of mid- to high-frequency optical modes
characterized by flat phonon dispersion and large frequency gaps. This is quantitatively demonstrated in Figs.~\ref{fig:figure5}(c) and S6(c), which present the frequency-resolved cumulative $\kappa_{\mathrm{c}}$ spectrum, its differential contribution, and atom-projected phonon density of states (DOS) at 300~K. Both Rb$_6$Re$_6$S$_8$I$_8$ and Cs$_6$Re$_6$S$_8$I$_8$ exhibit remarkably small $\kappa_{\mathrm{c}}$ values of 0.02 and 0.018~W$\cdot$m$^{-1}$$\cdot$K$^{-1}$, respectively, constituting only 12\% and 10\% of their total lattice thermal conductivity.
The primary $\kappa_{\mathrm{c}}$ contributions in Rb$_6$Re$_6$S$_8$I$_8$ originate from phonon modes below 2.5~THz, as evidenced by three distinct spectral peaks. To elucidate their origins, Fig.~\ref{fig:figure5}(d) maps the pairwise frequency combinations ($\omega_s$, $\omega_{s'}$)
contributing to $\kappa_{\mathrm{c}}$. The three peaks respectively arise from interactions between: (i) acoustic and the lowest optical branches, (ii) acoustic and low-lying optical branches, and (iii) low-lying optical branches themselves. DOS analysis reveals these spectral features
predominantly stem from vibrations involving Rb and I atoms. Notably, high-frequency optical modes above 11~THz exhibit non-negligible $\kappa_{\mathrm{c}}$ contributions due to their dispersive phonon band structures and strong mode hybridization. While Cs$_6$Re$_6$S$_8$I$_8$ shows similar $\kappa_{\mathrm{c}}$ characteristics, so it will not be discussed repeatedly.

\begin{figure}[ht]
	\centering
	\includegraphics[width=\textwidth,height=\textheight,keepaspectratio]{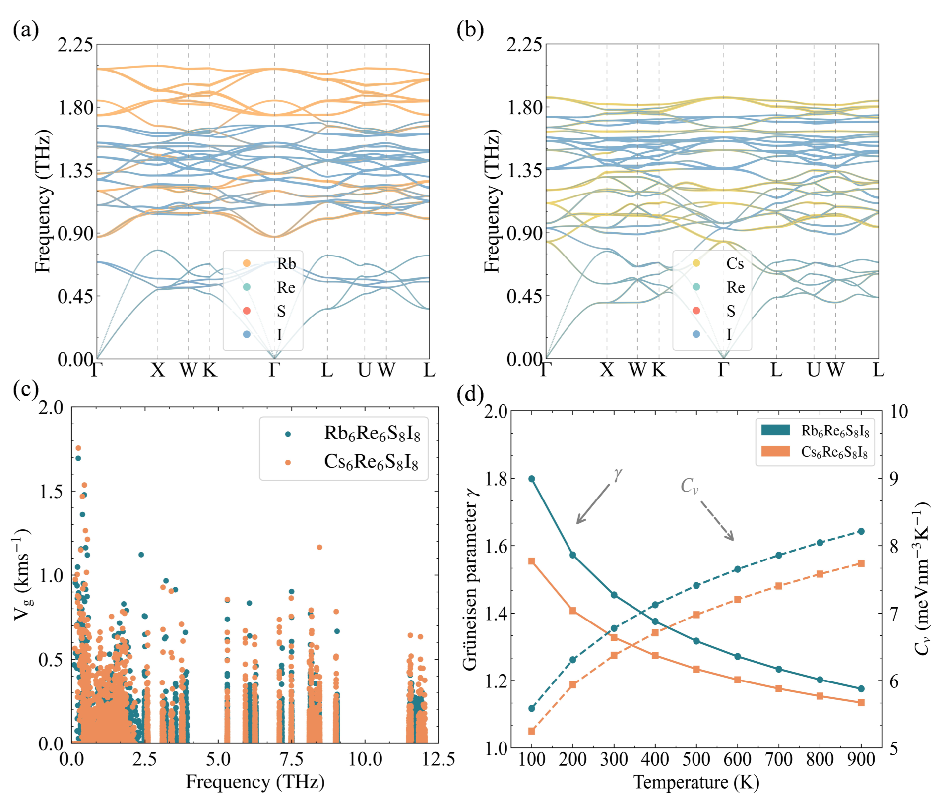}
	\caption{The low-frequency atomic projected phonon dispersion of (a) Rb$_6$Re$_6$S$_8$I$_8$, and (b) Cs$_6$Re$_6$S$_8$I$_8$ at 300 K. (c) The phonon group velocities as functions of thephonon frequency for X$_6$Re$_6$S$_8$I$_8$ (X = Rb, Cs) at 300 K. (d) The gruneisen parameter ($\lvert \gamma \rvert$) and the heat capacity (C$_v$) for X$_6$Re$_6$S$_8$I$_8$ (X = Rb, Cs) at 300 K.}
	\label{fig:figure6}
\end{figure}

A notable thermal transport anomaly emerges in X$_6$Re$_6$S$_8$I$_8$: Despite heavier atomic mass, Cs$_6$Re$_6$S$_8$I$_8$ exhibits lower lattice thermal conductivity than Rb$_6$Re$_6$S$_8$I$_8$. Mass perturbation tests reveal negligible $\rm \kappa_L$ change ($\rm \Delta\kappa_L \approx 0.001$ Wm$^{-1}$K$^{-1}$) when replacing Rb mass with Cs in Rb$_6$Re$_6$S$_8$I$_8$, indicating dominant interatomic interaction effects. Figure~\ref{fig:figure2}(d) reveals minimal changes in $\Phi_2$(Re-S) and $\Phi_2$(Re-I1) from Rb$_6$Re$_6$S$_8$I$_8$ to Cs$_6$Re$_6$S$_8$I$_8$, while $\Phi_2$(Re-I2) decreases by 9.7\%. In contrast, $\Phi_2$(Cs-I1) and $\Phi_2$(Cs-I2) increase by 16.5\% and 23.6\% compared to $\Phi_2$(Rb-I1) and $\Phi_2$(Rb-I2), respectively. As Re-I1 interactions mainly affect mid-frequency
optical branches with negligible $\rm \kappa_L$ contribution, the anomaly originates from enhanced X-I interactions. Bader charge analysis confirms stronger ionic bonding in Cs$_6$Re$_6$S$_8$I$_8$, with greater charge transfer between Cs and I2. X-I interactions predominantly modify acoustic and low-frequency optical branches (Fig.~\ref{fig:figure6}(a,b)). Rb$_6$Re$_6$S$_8$I$_8$ displays a small bandgap near 0.8~THz, flattening the lowest three optical branches and reducing phonon group velocities below
1~THz (Fig.~\ref{fig:figure6}(c)). Substituting $\Phi_2$(Cs-I) in Cs$_6$Re$_6$S$_8$I$_8$ with $\Phi_2$(Rb-I) softens phonon spectra (Fig.~S7). Weaker Rb-I interactions enhance anharmonicity, evidenced by higher $C_v$ and $\lvert \gamma \rvert$
values in Rb$_6$Re$_6$S$_8$I$_8$ compared to Cs$_6$Re$_6$S$_8$I$_8$ at identical temperatures (Fig.~\ref{fig:figure6}(d)).

\begin{figure}[ht]
	\centering
	\includegraphics[width=0.8\textwidth,keepaspectratio]{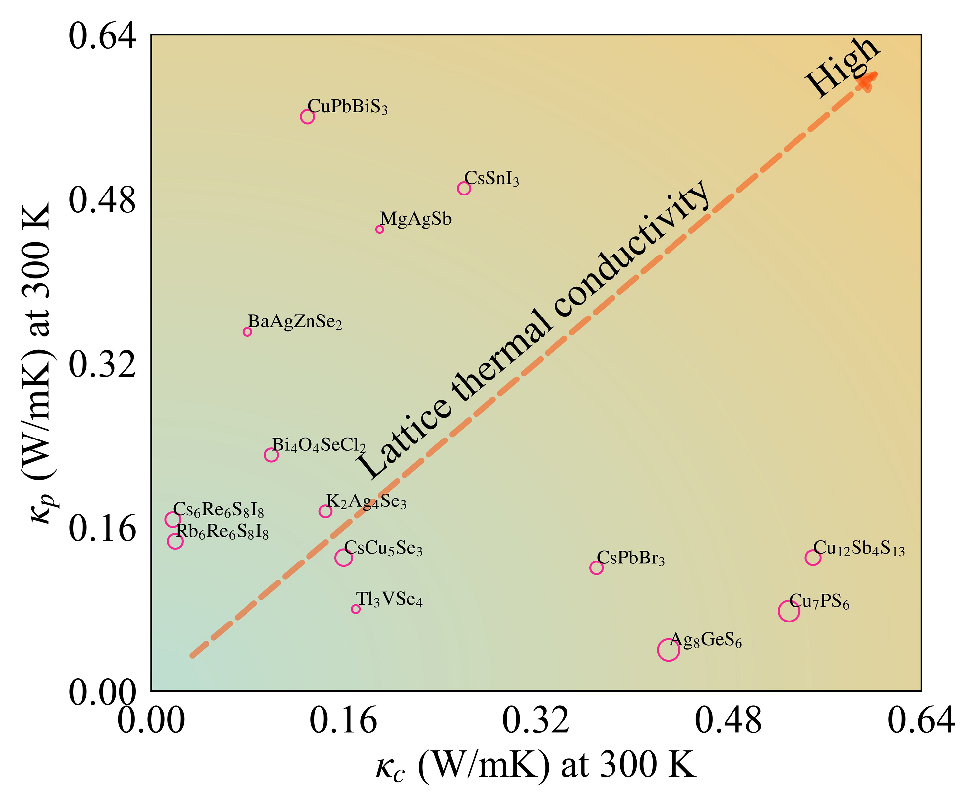}
	\caption{Potential pathways for pushing $\rm \kappa$ to its lower limit in inorganic materials. The dashed red line represents the current protocol to find materials with lower thermal conductivity. Lattice thermal conductivities are extracted from  previous studies based on unified theory \cite{zeng2024pushing,xia2020microscopic,simoncelli2019unified,tong2023glass,xie2023microscopic,shen2024amorphous,jain2020multichannel,li2024ultralow,xie2024anharmonic,pandey2022origin,li2025long,li2024crystal}.}
	\label{fig:figure7}
\end{figure}

The coexistence of ultralow particle-like $\kappa_{\mathrm{p}}$ and wave-like $\kappa_{\mathrm{c}}$ in X$_6$Re$_6$S$_8$I$_8$ represents a rare phenomenon in crystalline materials. While complex crystal structures generally exhibit suppressed $\kappa_{\mathrm{p}}$ through enhanced
phonon scattering (due to increased number of atoms in PC and Brillouin zone folding effects), excessive structural complexity may paradoxically enhance $\kappa_{\mathrm{c}}$. For instance, skutterudite YbFe$_4$Sb$_{12}$ (17 atoms in PC)\cite{di2023crossover} shows $\kappa_{\mathrm{p}}$ $=$ 0.83~W\,m$^{-1}$\,K$^{-1}$, and Bi$_4$O$_4$SeCl$_{2}$ (22 atoms in PC)\cite{tong2023glass} achieves $\kappa_{\mathrm{p}}$ $=$ 0.23~W\,m$^{-1}$\,K$^{-1}$. However, further
increasing crystal complexity (e.g., Cu$_7$PS$_6$ with 56 atoms/primitive cell \cite{PhysRevB.111.064307}) reduces the average phonon band spacing $\Delta\omega_{\mathrm{avg}}$, leading to non-negligible $\kappa_{\mathrm{c}}$ contributions exceeding 87\%. Figure~\ref{fig:figure7} maps the room-temperature $\kappa_{\mathrm{p}}$ and $\kappa_{\mathrm{c}}$ values against primitive cell atom counts for various materials. Although X$_6$Re$_6$S$_8$I$_8$ (28 atoms in PC) does not exhibit the lowest $\kappa_{\mathrm{p}}$, its unique
combination of high symmetry , moderate anharmonicity, and distinctive phonon flat bands which arising from bonding heterogeneity and guest-host architecture, enables unprecedented suppression of
$\kappa_{\mathrm{c}}$ ($<$ 0.02~~W\,m$^{-1}$\,K$^{-1}$). This contrasts with systems like AgTlI$_2$\cite{zeng2024pushing} where balanced $\kappa_{\mathrm{p}}$/$\kappa_{\mathrm{c}}$ contributions ($\sim$50\% each) yield $\kappa_{\mathrm{L}} = 0.25$~W$\cdot$m$^{-1}$$\cdot$K$^{-1}$, demonstrating the critical role of selective phonon engineering in achieving ultralow thermal conductivity.

\section{Conclusion}
In conclusion, based on the unified theory of lattice thermal transport, we have demonstrated that the superionic crystals Rb$_6$Re$_6$S$_8$I$_8$ and Cs$_6$Re$_6$S$_8$I$_8$ achieve ultralow lattice thermal conductivity $\kappa_{\mathrm{L}}$ of 0.17 and 0.19 ~W\,m$^{-1}$\,K$^{-1}$ at 300~K, respectively, with weak temperature dependence. This exceptional behavior stems from their unique hierarchical architecture: strongly bonded [Re$_6$S$_8$I$_6$]$^{4-}$ clusters interconnected via weak ionic interactions between X$^+$ (X = Rb, Cs) and I$^-$ ions. The relative vibrations between clusters generate extensive optical phonon flat bands, suppressing coherent thermal transport ($\kappa_{\mathrm{c}} \approx 0.02$~W\,m$^{-1}$\,K$^{-1}$). Concurrently, the soft X$^+$-I$^-$ interactions depress both low-frequency acoustic branches and optical modes, resulting in diminished particle-like conductivity ($\kappa_{\mathrm{p}} = 0.15$ and 0.17~W\,m$^{-1}$\,K$^{-1}$). Notably, Rb$_6$Re$_6$S$_8$I$_8$ exhibits anomalous thermal transport characteristics despite its lower atomic mass compared to the Cs analogue. The weaker Rb$^+$-I$^-$ coupling enhances lattice anharmonicity, reduces average phonon group velocities, and introduces a phonon band gap that further restricts dispersion in low-lying optical branches. These synergistic effects collectively yield the record-low $\kappa_{\mathrm{L}}$ in Rb$_6$Re$_6$S$_8$I$_8$. Our findings establish a new paradigm for phonon engineering through cluster-based anharmonicity tuning, providing a roadmap to develop ultralow $\kappa_{\mathrm{L}}$ materials for thermoelectric and thermal barrier applications.

\begin{acknowledgments}
This research was supported by the National Natural Science Foundation of China under Grants No. 12204402, 12304038, and the Scientific Research Foundation for Doctor of Jiangxi University of Science and Technology (jxxjbs18044); the Science and Technology Plan Project of Yingtan (202412-18904); Jiangxi Key Laboratory of Advanced Copper-based Materials (2024SSY05021).
\end{acknowledgments}

\bibliographystyle{unsrt}



\end{document}